\documentclass[%
 reprint,
]{revtex4-2}

\usepackage{graphicx}
\usepackage{dcolumn}
\usepackage{bm}

\begin{document}

\preprint{APS/123-QED}

\title{Local distortion driven magnetic phase switching in pyrochlore Yb$_2$(Ti$_{1-x}$Sn$_x$)$_2$O$_7$}

\author{Yuanpeng Zhang$^{1}$, Zhiling Dun$^{2}$, Yunqi Cai$^{2}$,  Chengkun Xing$^{3}$, Qi Cui$^{2}$,
    Naveen Kumar Chogondahalli Muniraju$^{4,5,6}$, Qiang Zhang$^{1}$, Yongqing Li$^{2}$, Jinguang Cheng$^{2,\dag *}$, Haidong Zhou$^{3,\ddagger}$}
 \email{$^{\dag}$jgcheng@iphy.ac.cn\\$^{\ddagger}$hzhou10@utk.edu}
\affiliation{%
 \vspace{0.3cm}
 $^1$Neutron Scattering Division, Oak Ridge National Laboratory, Oak Ridge, Tennessee 37831, USA\\
 $^2$Institute of Physics, Chinese Academy of Sciences, Beijing 100190, China\\
 $^3$Department of Physics and Astronomy, University of Tennessee, Knoxville, TN 37996, USA\\
 $^4$Forschungszentrum Jülich GmbH, Jülich Centre for Neutron Science (JCNS), Forschungszentrum Jülich, D-52425 Jülich, Germany\\
 $^5$Chemical and Engineering Materials Division, Oak Ridge National Laboratory, Oak Ridge, Tennessee 37831, USA\\
 $^6$Institute of Physics, Bijenička cesta 46, 10000, Zagreb, Croatia
}%

\date{\today}

\begin{abstract}
While it is commonly accepted that the disorder induced by magnetic ion doping in quantum magnets usually generates a rugged free-energy landscape resulting in slow or glassy spin dynamics, the disorder/distortion effects associated with non-magnetic ion sites doping are still illusive. Here, using AC susceptibility measurements, we show that the mixture of Sn/Ti on the non-magnetic ion sites of pyrochlore Yb$_2$(Ti$_{1-x}$Sn$_x$)$_2$O$_7$ induces an antiferromagnetic ground state despite both parent compounds, Yb$_2$Ti$_{2}$O$_7$, and Yb$_2$Sn$_{2}$O$_7$, order ferromagnetically. Local structure studies through neutron total scattering reveals the local distortion in the non-magnetic ion sites and its strong correlation with the magnetic phase switching. Our study, for the first time, demonstrates the local distortion as induced by the non-magnetic ion site mixture could be a new path to achieve magnetic phase switching, which has been traditionally obtained by external stimuli such as temperature, magnetic field, pressure, strain, light etc.
\end{abstract}

\maketitle

The complex interplay between various degree of freedom in frustrated quantum systems enables rich physics involving complex ground states and exotic excitations \cite{RevModPhys.82.53, PhysRevB.96.174418}. The balancing between those weak interaction terms in the Hamiltonian (long-range dipolar interactions, exchange interaction beyond the nearest neighboring, site disorder, lattice distortions, etc.) and the frustration could determine the coordination in the phase diagram \cite{rossphdthesis, PhysRevLett.103.227202, PhysRevB.98.094412, PhysRevLett.109.097205}. Exploring such a complex interplay thus becomes critical in terms of understanding the formation of those exotic states such as the quantum spin liquid (QSL) state where the long range magnetic ordering is completely suppressed by frustration and the quantum fluctuation yields strong dynamics even at zero temperature. In practice, the QSL state of matter (e.g., the non-Abelian anyons in 2D QSL) hosts promising application in error-proof quantum computation \cite{Banerjee2018, RevModPhys.80.1083, PhysRevLett.59.1776} and also is believed to be closely related to high temperature superconductivity \cite{Chamorro2021, Balents2010, doi:10.1126/science.235.4793.1196}.

For the study of frustrated quantum systems, of critical concern is the underlying lattice characteristics, such as the chemical order/disorder, lattice/sub-lattice distortion, etc. For example, the impact of various types of disorder has been attracting notable attention -- this is especially true for the disorder effect with respect to the magnetic species, as they are directly (predominantly through near neighbor interaction) associated with the magnetic Hamiltonian. Depending on the level of disorder, it can either perturb spontaneous symmetry breaking \cite{Villain1979} or promote magnetic ordering and break a continuous degeneracy via an order by disorder mechanism \cite{PhysRevLett.62.2056}. Specifically concerning the pyrochlore systems, on which the current report is to focus, there have been extensive studies over the magnetic species relevant disorder effects, including those diluted \cite{PhysRevLett.92.107204, Sheetal_2020, PhysRevX.9.031047, PhysRevLett.129.037204, PhysRevX.11.011034} or stuffed \cite{PhysRevB.86.174424, Bowman2019, Lau2006, Gardner_2011} pyrochlores. Following similar pathway, there have been some recently emerging studies on the disorder effect of the non-magnetic-only species on the non-magnetic site. Several celebrated examples recently on this topic are YbMgGaO$_4$ (YMGO) \cite{PhysRevLett.117.097201, PhysRevLett.122.137201, Shen2016, Paddison2017, Shen2018}, Sr$_3$CuTa$_2$O$_9$ (SCTO) \cite{sana2023spinliquidlike}, Sr$_2$Cu (Te$_{1-x}$W$_x$)O$_6$ (SCTWO) \cite{PhysRevLett.126.037201, PhysRevB.105.184410, PhysRevB.107.L020407}, SrLaCuSbO$_6$ (SLCSO) \& SrLaCuNbO$_6$ (SLCNO) \cite{PhysRevB.105.054414}, and Lu$_3$Sb$_3$Mn$_2$O$_{14}$ (LSMO) \cite{PhysRevB.107.214404}. As compared to the magnetic species involved chemical disorder, first, the non-magnetic-only disorder has to take effect through the magnetic interactions beyond the nearest-neighbor. For example, in SCTWO, substituting W for Te alters the magnetic interactions from the strong nearest-neighbor type to the strong next-nearest-neighbor type \cite{PhysRevB.105.184410}, resulting in strong exchange interaction disorder that is absent in parent compounds Sr$_2$CuTeO$_6$ \cite{PhysRevB.93.054426} and Sr$_2$CuWO$_6$ \cite{Vasala_2014, PhysRevB.89.134419}. Theoretically, the scenario could be described as the random-singlet (RS) state \cite{PhysRevB.107.L020407, PhysRevLett.126.037201, PhysRevX.8.031028}, in which the randomness in a quantum magnet can induce spin-singlet dimers of varying strengths with a spatially random manner and therefore account for the spin liquid like behaviors. Second, for non-magnetic-only order/disorder systems, it is relatively easier to separate out the lattice distortion effect from the chemical order/disorder effect, especially for those systems in which the magnetic interactions beyond the nearest neighbor are non-critical.

Equally important as the order/disorder effect, the local lattice distortion also plays manifold critical roles in promoting different ground states and quantum excitations. Both the crystal field environment and the delicate balance between the anisotropic exchanges could be tuned by the local distortions \cite{RevModPhys.88.041002, Yadav2016, Balz2016}, and such an interplay could potentially induce accidental degeneracy in the vicinity of the phase boundary and thus could lead to the emergence of a QSL \cite{Kermarrec2017}. Another typical and intricate scenario involving local distortion is for those systems presenting strong spin-orbital coupling -- on one side, the local distortion is potentially destroying the degeneracy of energy levels and thus suppressing the quantum fluctuation that is necessary for the formation of QSL states. On the other, the local symmetry breaking may lead to orbital quenching and therefore suppress the spin-orbital coupling, which is then beneficial for the QSL states formation, since the spin-orbital coupling is susceptible of lifting the degeneracy of energy levels and thus is detrimental for the QSL states. As such, several typical examples were showing the dominant effect of suppressing the quantum fluctuations and its winning over the orbital quenching effect to give exotic quantum states were reported for the NiRh$_2$O4 system \cite{PhysRevMaterials.2.034404, PhysRevLett.120.057201, PhysRevB.100.140408, PhysRevB.96.020412}. Such impacts of local distortion, along with other potential effects such as the formation of local dimer or trimer clusters \cite{PhysRevLett.124.167203, Kelly2019, PhysRevLett.111.217201}, induction of fast spin fluctuations \cite{PhysRevB.84.184409}, reduction of the effective dimension of magnetic coupling \cite{Zhang2018}, and among others, infers the importance of directly probing the local distortion and constructing a clear picture of the interplay with magnetic coupling.

With this regard, extensive existing studies focus on the impact of local distortion on the scheme of exchange interactions, and among other aspects such as electron localization and charge density wave. Typically for QSL candidate systems, various studies reported the direct probe of the local environment and its link to the exotic QSL states, using NMR \cite{PhysRevMaterials.2.044406, PhysRevB.101.214402}, Electron Spin Resonance (ESR) \cite{PhysRevLett.118.017202}, X-ray Absorption Spectroscopy (XAS) \cite{PhysRevMaterials.4.124406}, and X-ray/neutron total scattering \cite{PhysRevB.84.064117, bozin2023hightemperature}. We realized enormous such research in the herbertsmithite and barlowite QSL candidate systems. However, for the pyrochlore systems, direct experimental probing of the local structural variation and its link to the magnetic coupling is still lacking, though, excessive theoretical work has provided clear indication for the impact of local structural distortion, via the spin-lattice coupling effect, upon the magnetic states and spin ordering \cite{PhysRevB.99.144406, PhysRevLett.107.047204, PhysRevB.93.220407, PhysRevLett.112.167203, PhysRevLett.118.087203, PhysRevLett.110.127206, PhysRevLett.90.147204}. To the best of our knowledge, there is only limited experimental work focusing on the spin-lattice coupling effect in pyrochlore systems from the local perspective, like the report by P. M. Thygesen, {\em et al.} demonstrating the local orbital dimerization of Jahn-Teller active Mo$^{4+}$ ions instead of random compositional or site disorder drives the spin-glass state in Y$_2$Mo$_2$O$_7$ \cite{PhysRevLett.118.067201}. More relevant experimental efforts, though still limited, were mainly from the average long-range structure perspective \cite{PhysRevB.89.064409, PhysRevB.103.214418, PhysRevB.93.134426, PhysRevB.86.174424}. However, while the local distortion could be extracted from conventional Bragg diffraction, the local and short-range probe could provide unique pathway and different angle to such explorations, as is already demonstrated by the work on the herbertsmithite and barlowite QSL candidate systems \cite{PhysRevMaterials.2.044406, PhysRevB.101.214402, PhysRevLett.118.017202, PhysRevMaterials.4.124406, PhysRevB.84.064117, bozin2023hightemperature}. In this report, we were trying to utilize the neutron total scattering measurement, which incorporates both the Bragg peaks and the diffuse scattering signal, to study the spin-lattice coupling in the Sn-doped Yb$_2$Ti$_{2}$O$_7$ (YTSO). For such a system, it is believed only the nearest neighbor coupling is dominant in the magnetic Hamiltonian \cite{PhysRevB.100.104423, PhysRevX.1.021002, PhysRevLett.109.097205, PhysRevB.87.184423, PhysRevLett.106.187202}. Meanwhile, the doping in our studied YTSO system is only on the non-magnetic site, which infers the chemical disorder effect in the YTSO system could be singled out from the potentially existing distortion effect. Through our modeling for the neutron total scattering data together with magnetic susceptibility measurements, a magnetic phase switching could be clearly identified, as proposed to be induced by the local distortion associated with the Sn-doping on the non-magnetic Ti sites.

\begin{figure}[hb]
\includegraphics[width=8.6cm]{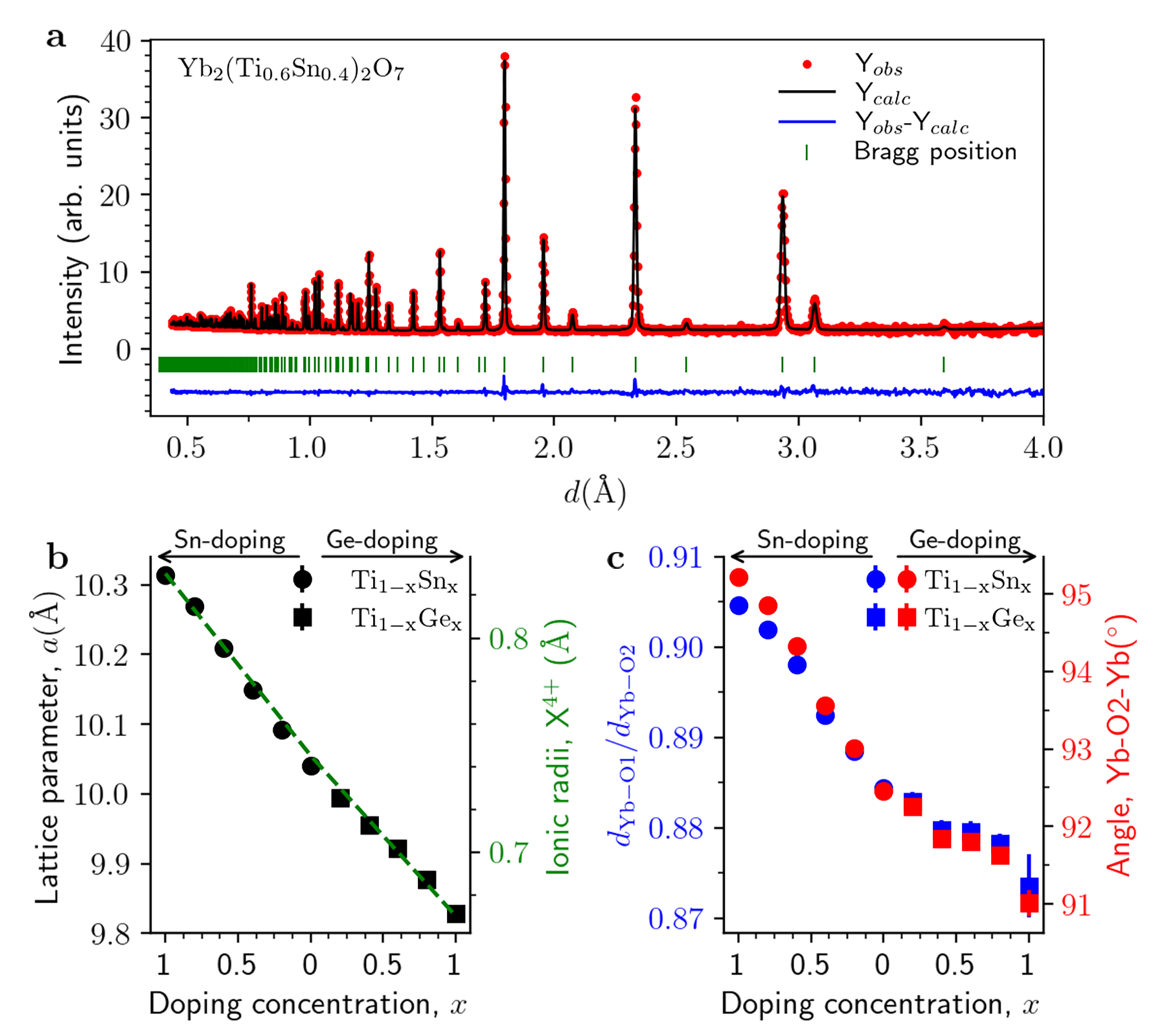}
\caption{\label{fig1}{\bf Average structure of Yb$_2$(Ti$_{1-x}$Sn$_{x}$)$_2$O$_7$ and Yb$_2$(Ti$_{1-x}$Ge$_{x}$)$_2$O$_7$}. {\bf a.} Neutron powder diffraction pattern (red circles) for Yb$_2$(Ti$_{0.6}$Sn$_{0.4}$)$_2$O$_7$ measured at 300 K with the central wavelength of 1.5 \AA \ on POWGEN diffractometer. The solid black line is the Rietveld refinement using FullProf \cite{RODRIGUEZCARVAJAL199355}. Solid blue line at the bottom of the panel shows the difference curve. The Bragg peaks are marked with green markers. {\bf b.}  The doping concentration dependence of the lattice parameter. {\bf c.}  The doping level dependence of the ratio between the two different Yb-O bond lengths and the Yb-O2-Yb angle.}
\end{figure}

\textbf{Neutron diffraction.}
For comparison, we synthesized both Yb$_2$(Ti$_{1-x}$Sn$_{x}$)$_2$O$_7$ and Yb$_2$(Ti$_{1-x}$Ge$_{x}$)$_2$O$_7$ samples and used neutron powder diffraction (NPD) to characterize their lattice structures.  Fig. 1(a) shows the refinement for the NPD data of Yb$_2$(Ti$_{0.6}$Sn$_{0.4}$)$_2$O$_7$  measured at room temperature using the POWGEN diffractometer. The data could be well fitted by the $Fd\bar{3}m$ pyrochlore structure. The NPD data for several other Sn and Ge doped samples was also refined (not shown here), which all exhibits pure pyrochlore structure. As summarized in Fig. 1(c), the lattice parameter $a$  decreases from Yb$_2$Sn$_2$O$_7$ to Yb$_2$Ti$_2$O$_7$ and then Yb$_2$Ge$_2$O$_7$ for all doped samples. This is reasonable since the lattice parameter is strongly related to the ironic radius of the (Sn/Ti/Ge) site, and therefore the Sn sample has the largest lattice parameter, Ti sample the second, and Ge sample the smallest. We further used $\rho = d_\textrm{Yb-O2}/d_\textrm{Yb-O1}$ and the Yb-O2-Yb angle to characterize the axial distortion of the YbO$_8$ polyhedra, here $d_\textrm{Yb-O1}$ represents the bond length for the 6 longer Yb-O1 bonds in the plane perpendicular to the $\langle111\rangle$ axis and $d_\textrm{Yb-O2}$ represents the bond length for the two shorter Yb-O2 bonds along the $\langle111\rangle$ axis. As shown in Fig. 1(d), again, both of them decrease linearly, without abrupt change.

\begin{figure*}[!hbt]
\includegraphics[width=17.2cm]{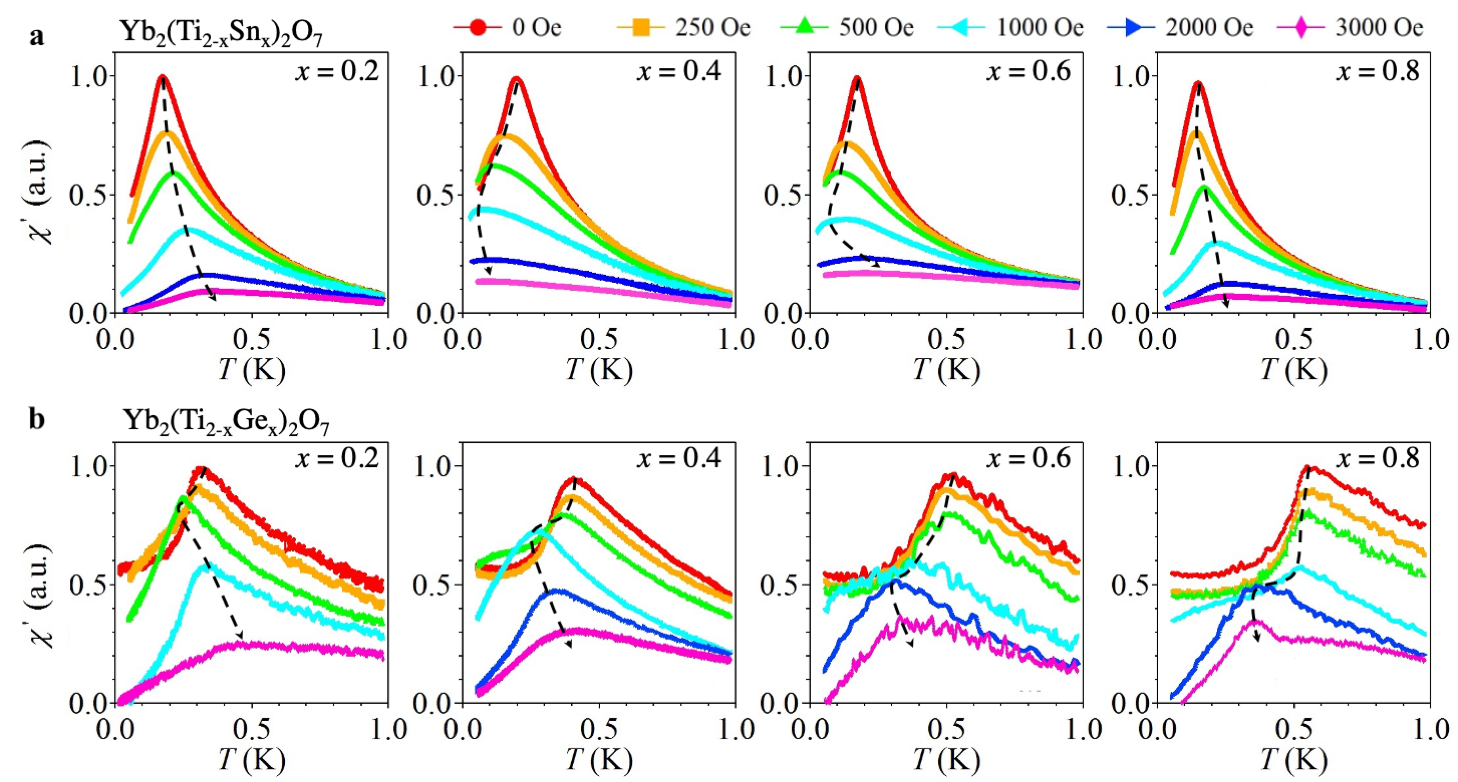}
\caption{\label{fig2}{\bf AC susceptibility}. {\bf a.} Real part of AC susceptibility measured under different DC magnetic fields in the arbitrary unit (a.u.) for Yb$_2$(Ti$_{1-x}$Sn$_{x}$)$_2$O$_7$ with $x$ = 0.2, 0.4, 0.6, and 0.8. {\bf b.} Similar data for Yb$_2$(Ti$_{1-x}$Ge$_{x}$)$_2$O$_7$  with $x$ = 0.2, 0.4, 0.6, and 0.8. The used AC frequency is 3317 Hz for the Sn-doped system and 331 Hz for the Ge-doped system with a magnitude of 5 Oe.  Dashed arrows indicate the evolution of the peak's position with increasing DC fields. }
\end{figure*}

\textbf{AC susceptibility.}
Fig. 2 shows the AC susceptibility measured at different DC magnetic fields for Yb$_2$(Ti$_{1-x}$Sn$_{x}$)$_2$O$_7$ and Yb$_2$(Ti$_{1-x}$Ge$_{x}$)$_2$O$_7$. For all samples, the data at zero field exhibits a peak, which represents the long range magnetic ordering at $T$*, and such a peak shows an obvious shift under applied DC fields. As demonstrated in Ref. \cite{PhysRevB.89.064401}, the field dependence of AC susceptibility can be used as a convenient tool to identify the nature of a long-range magnetic ordering, i.e., the ferromagnetic (FM) ordering temperature will shift to higher temperatures with increasing DC field due to the contribution of domain magnetization while the antiferromagnetic (AFM) ordering temperature will show an opposite DC field dependence.

The field dependence of $T$* for each doping level was summarized in Fig. 3(a). The data shows (i) for Yb$_2$Ti$_2$O$_7$ and Yb$_2$Sn$_2$O$_7$, the $T$* increases with increasing field, which is consistent with the fact that both samples have a splayed ferromagnetic (SF) ground state; (ii) for Yb$_2$Ge$_2$O$_7$, the $T$* decreases with increasing field, which is consistent with its AFM ground state \cite{PhysRevB.92.140407,PhysRevB.93.104405,PhysRevB.89.064401}; (iii) for all Sn- and Ge-doped samples except Yb$_2$(Ti$_{0.8}$Sn$_{0.2}$)$_2$O$_7$, the $T$* decreases first with increasing field and then increases while the field surpasses a critical value $H_{\text{c}}$. This indicates that as soon as Sn and Ge are doped, certain volume of AFM phase is introduced. This AFM order should be in long range nature since it dominates the bulk magnetism at low fields. With $H$ $>$ $H_{\text{c}}$, the sample comes back to ferromagnetic or is fully polarized; (iv) for Yb$_2$(Ti$_{0.8}$Sn$_{0.2}$)$_2$O$_7$, it has ferromagnetic ground state since its $T$* monotonically increases with increasing field.

\textbf{Magnetic Phase Diagram.}
Accordingly, a magnetic phase diagram of $T$* and $H_{\text{c}}$ for Yb$_2$(Ti$_{1-x}$Sn$_{x}$)$_2$O$_7$ and Yb$_2$(Ti$_{1-x}$Ge$_{x}$)$_2$O$_7$ is summarized in Fig. 3(b). For Ge-doped samples, both $T$* and $H_{\text{c}}$ monotonically increase with increasing Ge-doping level. On the other hand, for Sn-doped samples, (i) while the $T$* generally decreases with increasing Sn-doping level, it exhibits a dome around  $x$ = 0.5;  (ii) the $H_{\text{c}}$ first increases with increasing Sn-doping level, peaks at $x$ = 0.5, and thereafter decreases.

\begin{figure}
\includegraphics[width=8.6cm]{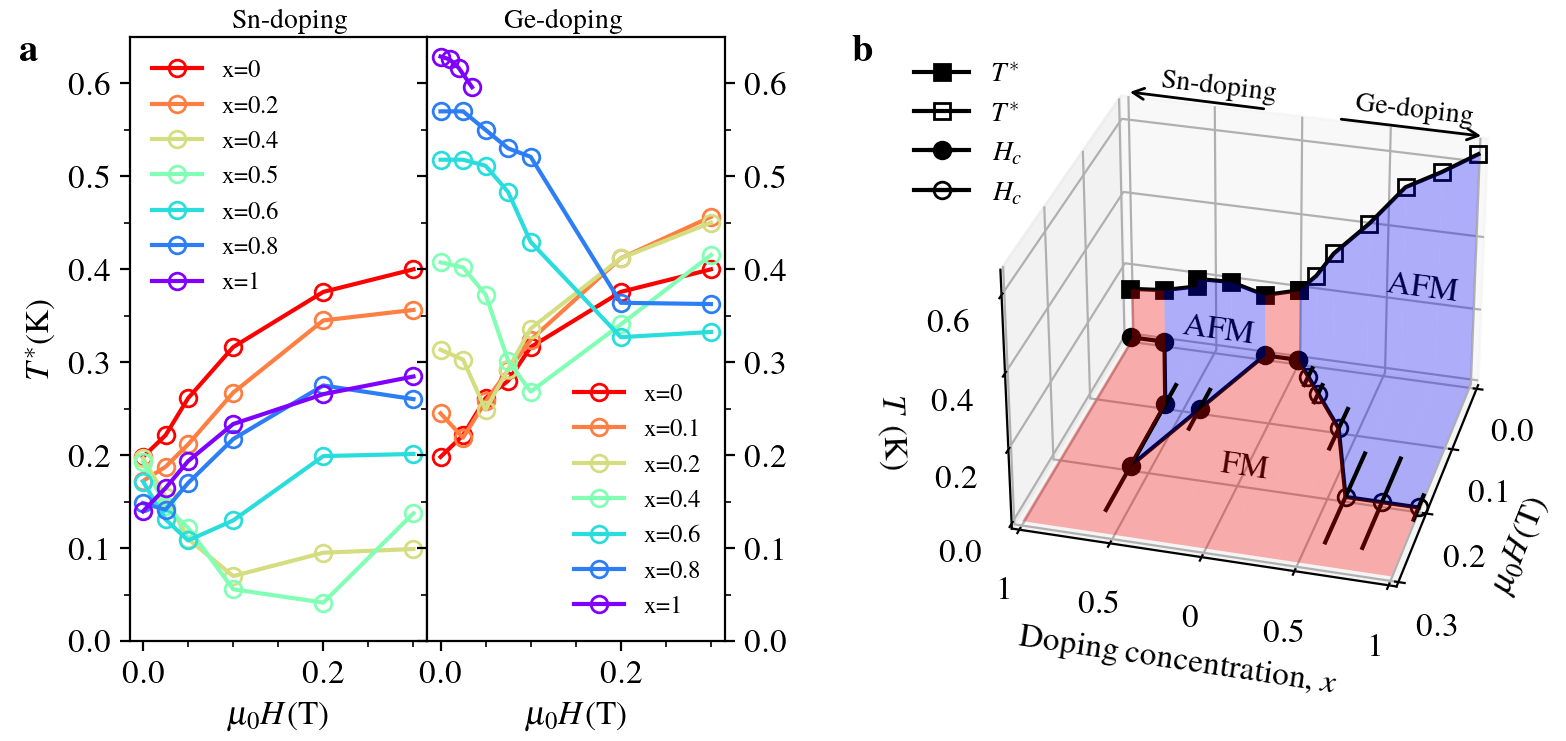}
\caption{\label{fig3}{\bf Magnetic phase diagram.} {\bf a.} The field dependence of the magnetic ordering temperature T$^{*}$ for Yb$_2$(Ti$_{1-x}$Sn$_{x}$)$_2$O$_7$ and Yb$_2$(Ti$_{1-x}$Ge$_{x}$)$_2$O$_7$ with different doping concentrations. {\bf b.} Magnetic phase diagram as a function of the doping concentration, $x$, temperature, $T$, and the external DC field, $\mu_0H$. The red and blue regions represent the FM and AFM phases, respectively. Data points for Yb$_2$Ti$_2$O$_7$, Yb$_2$Sn$_2$O$_7$, and Yb$_2$Ge$_2$O$_7$ are from Dun et al. \cite{PhysRevB.89.064401}.}
\end{figure}

The evolution of $T$* and $H_{\text{c}}$ in the Ge-doped samples is expected. The magnetic ground states of Yb-pyrochlores are determined by the ratio among the anisotropic exchange interactions \cite{PhysRevB.92.140407,PhysRevB.88.144402,PhysRevX.7.041057}. Unlike Yb$_2$Ti$_2$O$_7$, Yb$_2$Ge$_2$O$_7$ orders antiferromagnetically in the $\Gamma_5$ manifold \cite{PhysRevB.93.104405}. From the point view of chemical pressure effect, with increasing the Ge-doping level in Yb$_2$(Ti$_{1-x}$Ge$_{x}$)$_2$O$_7$, the lattice parameter decreases monotonically and gradually tunes the balance of anisotropic exchange interactions that drives the system towards the AFM $\Gamma_5$ phase from the SF phase \cite{PhysRevB.92.140407,PhysRevB.93.104405}. Alternatively, if we assume that the phase coexistence in Ge-doped samples is similar to that of Yb$_2$Ti$_2$O$_7$, then the $H_{\text{c}}$ could be scaled to the volume fraction of the AFM phase since the larger the $H_{\text{c}}$ is, the more difficult to polarize the system. Therefore, the evolution of $H_{\text{c}}$ in Fig. 3(b) means a monotonic increase in volume fraction of the AFM phase in Yb$_2$(Ti$_{1-x}$Ge$_{x}$)$_2$O$_7$, consistent with the SF and AFM phases in Yb$_2$Ti$_2$O$_7$ and Yb$_2$Ge$_2$O$_7$, respectively.

However, although we cannot determinate whether it is the $\psi_2$ or $\psi_3$ phase of the $\Gamma_5$ manifold, the appearance of the long range AFM order in Yb$_2$(Ti$_{1-x}$Sn$_{x}$)$_2$O$_7$ is surprising. Since both Yb$_2$Ti$_2$O$_7$ and Yb$_2$Sn$_2$O$_7$ have the SF ground state, an AFM ground state should not be expected for Sn-doped samples from the view of chemical pressure effects. Even if there is still magnetic phase coexistence \cite{doi:10.1073/pnas.2008791117}, it is puzzling to observe this non-monotonic change of the volume fraction of AFM phase (or, $H_{\text{c}}$) in Yb$_2$(Ti$_{1-x}$Sn$_{x}$)$_2$O$_7$.

\begin{figure}[!hbt]
\includegraphics[width=8.6cm]{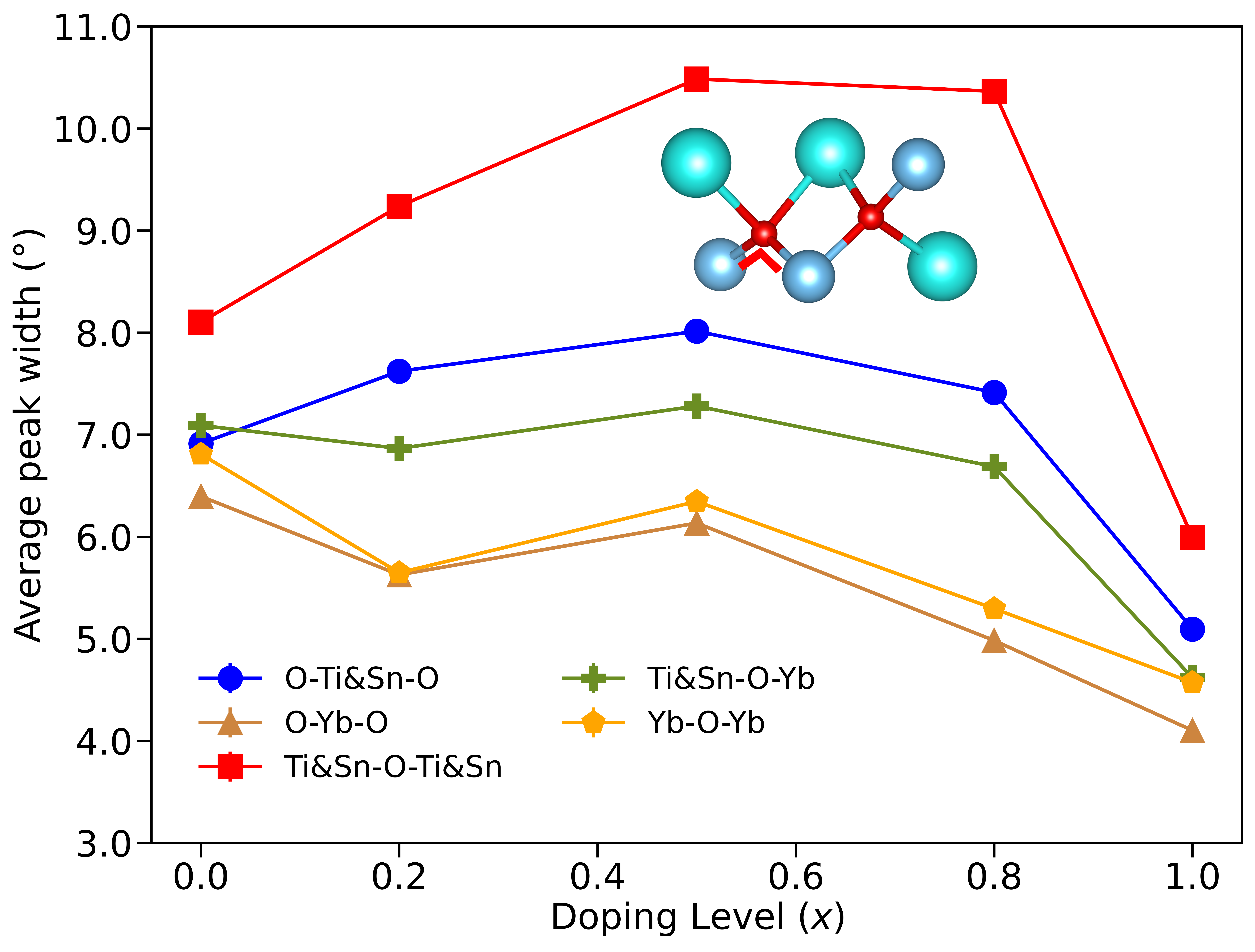}
\caption{\label{fig4}{\bf Local distortion as characterized by various local bond angles, as extracted from the RMC resulted configuration}. The inset is an illustration for the local geometry involving the critical triplet angle -- Ti\&Sn-O-Ti\&Sn. Yb $\rightarrow$ cyan, Ti\&Sn $\rightarrow$ blue, O $\rightarrow$ red. Error bar was estimated via running 10 repeated RMC modeling followed by the same bond angle calculation. The actual error bar is smaller than the symbol for all the data points as presented and here we just take the symbol size as the representative error bar level.}
\end{figure}

\textbf{Total Scattering.}
The total scattering signal contains both the Bragg peaks and the diffuse scattering contribution. The diffuse scattering part, which is usually taken as the background and thus subtracted off in conventional Bragg peaks analysis, in fact provides unique access to the local structure. Although both the Bragg peaks and total scattering data would yield information about local distortions, the former is based on {\em the distance between average positions} whereas the latter is based on {\em the average of distances ensemble}. Therefore, the local probe via total scattering data could provide access to the local structural variation that is inaccessible through the conventional Bragg analysis. Here, we collected neutron total scattering data on POWGEN for the series of Yb$_2$(Ti$_{1-x}$Sn$_{x}$)$_2$O$_7$ samples and performed the reverse Monte Carlo (RMC) \cite{Zhang:kc5118, Tucker_2007} modeling \cite{the_si1} to extract the local structure. A 10$\times$10$\times$10 supercell containing more than 100, 000 atoms could be constructed from the RMC modeling and the following statistical calculation could be performed to extract the key structural aspects. Typically, here we focused on the various bond angles involving both the Yb, Ti/Sn, and O atoms -- from the RMC resulted configuration, a statistical distribution of various triplet angles could be obtained \cite{the_si2}. Further, the width of the triplet angles distribution could be extracted through a Gaussian peak fitting, which indicates the dispersiveness of the bond angle and thus infers the significance of the local distortion. The results are presented in Fig. \ref{fig4}. No abrupt change as the function of the doping level could be observed except for the Ti\&Sn-O-Ti\&Sn triplet (see the inset of Fig. \ref{fig4}). This indicates the non-magnetic-site involved local distortion is enhanced as the result of the doping. Assuming the nearest neighbor magnetic interaction is dominant in the YTSO system, the result here infers the potential link between the magnetic phase switching and the local distortion. As of our knowledge, this is the first experimental evidence demonstrating the magnetic phase switching in pyrochlore systems that is potentially coupled with the {\em local structural distortion} beyond the compositional randomness or site disorder. Our finding here falls in line with various theoretical studies for the spin-lattice coupling in pyrochlore systems. For example, through many-body quantum-chemical calculations, N. Bogdanov {\em et al.} showed the magnetic interactions and ordering in Cd$_2$Os$_2$O$_7$ are crucially dependent on the local geometrical features \cite{PhysRevLett.118.087203}. H. Shinaoka, et al. showed the coupling of spin-glass transitions to local lattice distortions on pyrochlore lattices via Monte Carlo simulation \cite{PhysRevLett.107.047204}. Also using Monte Carlo simulation, K. Aoyama {\em et al.} revealed a lattice distortion induced spin ordering in the breathing pyrochlore lattice\cite{PhysRevB.99.144406}. Based on such theoretical results, we believe the non-magnetic site mixing in the YTSO system is imposing effect upon the magnetic coupling through the induced local lattice distortion, which then further onsets the magnetic phase switching at a certain level of mixing. According to a recent report by A. Scheie {\em et al.}, it was revealed that the QSL physics in Yb$_2$Ti$_2$O$_7$ is fundamentally the FM-AFM phase competition in the low temperature regime \cite{doi:10.1073/pnas.2008791117}. With this regard, the FM-AFM phase switching and its link to the local lattice distortion is of strong relevance and infers that the non-magnetic site doping and the accordingly induced local structural variation should be paid serious attention for the study of QSL physics in pyrochlore systems.

As a final remark, we want to emphasize that the conclusion we arrived at here about there being a strong correlation between the local distortion and magnetic phase switching is partly based on the assumption that the magnetic coupling beyond the nearest neighbor is not taking critical effect in the YTSO system -- in fact, such an assumption was indeed adopted in various previous reports as mentioned earlier \cite{PhysRevB.100.104423, PhysRevX.1.021002, PhysRevLett.109.097205, PhysRevB.87.184423, PhysRevLett.106.187202}. Without such an assumption, the induced randomness in the Hamiltonian as the result of chemical disorder and its effect upon the magnetic ordering cannot be rigorously excluded. As such, future studies are needed -- theoretically, magnetic interaction beyond the nearest neighbor needs to be inspected and its effect on the magnetic phase diagram should be studied.

\begin{acknowledgments}
We thank Martin Mourigal and Itamar Kimchi for helpful discussion.
J.G.C. is supported by the National Natural Science Foundation of China (12025408, 11874400, 11921004), the Key Research Program of Frontier Sciences of CAS ( QYZDB-SSW-SLH013), the CAS Interdisciplinary Innovation Team (JCTD-2019-01) and Lujiaxi international group funding of K. C. Wong Education Foundation (GJTD-2020-01). The work at the University of Tennessee is supported by the U. S. Department of Energy under grant No. DE-SC0020254. Part of the research conducted at SNS was sponsored by the Scientific User Facilities Division, Office of Basic Energy Sciences, US Department of Energy. The following funding is acknowledged: US Department of Energy, Office of Science (contract No. DE-AC05-00OR22725). This research used resources of the Computeand Data Environment for Science (CADES) at the Oak Ridge National Laboratory,which is supported by the Office of Science of the U.S. Department of Energy under Contract No. DE-AC05-00OR22725. This research used resources of the Compute and Data Environment for Science (CADES) at the Oak Ridge National Laboratory, which is supported by the Office of Science of the U.S. Department of Energy under Contract No. DE-AC05-00OR22725. This research used resources of the National Energy Research Scientific Computing Center (NERSC), a U.S. Department of Energy Office of Science User Facility located at Lawrence Berkeley National Laboratory, operated under Contract No. DE-AC02-05CH11231 using NERSC award ERCAP0024340. We thank Dr. Emil S. Bozin for the discussion about the local structure studies.
\end{acknowledgments}

\nocite{*}

\bibliography{ytso_refs}

\begin{thebibliography}{84}
\bibitem [{the({\natexlab{a}})}]{the_si1}%
  \BibitemOpen
  \href@noop {} {\bibinfo {title} {See {S}upplemental {M}aterial at [{URL} to
  be inserted] for the technical details about the reverse {M}onte {C}arlo
  modeling}} \BibitemShut {NoStop}%
\end{thebibliography}

\end{document}